\documentclass[namedreferences,optionalrh,solaromanenum]{spr-sola}
\usepackage{graphicx}                    
\usepackage{subfig}
\usepackage{amssymb}                    
\usepackage{color}                       
\usepackage{pdflscape}
\usepackage{longtable}
\usepackage{booktabs}
\usepackage{caption}
\usepackage{url}

\usepackage[dvipsnames]{xcolor}

\usepackage[
    colorlinks=true,
    linkcolor=blue,
    citecolor=blue,
    urlcolor=blue,
    pdfauthor={Dr. K. Sasikumar Raja},
    pdftitle={VELC Data Processing},
    pdfsubject={Solar Observations},
    pdfkeywords={VELC, Aditya-L1, solar corona}
]{hyperref}

\usepackage{rotating} 
\usepackage{array}
\usepackage{longtable}
\usepackage{supertabular}
\usepackage{booktabs}  

\chardef\us=`\_
\begin{document}

\begin{frontmatter}

\title{Relative Strengths of Fundamental and Harmonic Emissions of Solar Radio Type II Bursts}

%
\author[addressref={aff1},email={rishikesh.govind@iiap.res.in}]{\inits{Rishikesh G. }\fnm{Rishikesh G. }\snm{Jha}\orcid{0009-0001-0122-583X}}

\author[addressref={aff1},corref,email={sasikumar.raja@iiap.res.in}]{\inits{K. }\fnm{K. }\snm{Sasikumar Raja}\orcid{0000-0002-1192-1804}}

\author[addressref={aff1},email={ramesh@iiap.res.in}]{\inits{R. }\fnm{R. }\snm{Ramesh}\orcid{0000-0002-1192-1804}}

\author[addressref={aff1},email={kathir@iiap.res.in}]{\inits{C. }\fnm{C. }\snm{Kathiravan}\orcid{0000-0002-6126-8962}}

\author[addressref={aff2},email={christian.monstein@irsol.usi.ch}]{\inits{C. }\fnm{Christian }\snm{Monstein}\orcid{0000-0002-3178-363X}}

%
\runningauthor{Jha et al}
\runningtitle{Relative Strengths of Fundamental and Harmonic Emissions of Solar Radio Type II Bursts}

\address[id=aff1]{Indian Institute of Astrophysics, Koramangala-560 034, Bengaluru, India}

\address[id=aff2]{Istituto ricerche solari Aldo e Cele Dacco` (IRSOL), Universita` della Svizzera italiana, Locarno, Switzerland}
\begin{abstract}
Solar radio type II bursts are slow-drifting bursts that exhibit various distinct features such as Fundamental (F) and Harmonic (H) emissions, band-splitting, and discrete fine structures in the dynamic spectra. Observationally, it has been found that in some cases the F emission is stronger than the H emission, and vice versa. The reason for such behavior has not been thoroughly investigated. To investigate this, we studied 58 meter wave (20-500 MHz) type II solar radio bursts showing both F and H emissions, observed during the period from 13 June 2010 to 25 December 2024, using data obtained with the Compound Astronomical Low frequency Low cost Instrument for Spectroscopy and Transportable Observatory (CALLISTO) spectrometers at different locations and Gauribidanur LOw-frequency Solar Spectrograph (GLOSS). We examined the intensity ratios of the H ($I_H$) and F ($I_F$) emissions and analyzed their variation with heliographic longitude. We found that 14 out of 19 bursts originating from heliographic longitudes beyond $\pm75^\circ$ exhibited an $I_H/I_F$ ratio greater than unity. In contrast, 32 out of 39 bursts originating from longitudes within $\pm75^\circ$ showed a intensity ratio less than unity. From these results, we conclude that the relative strength of the F and H emissions can be influenced by refraction due to density gradient in the solar corona, directivity and viewing angle of the bursts.
\end{abstract}

%
\keywords{Solar radio bursts, Type II bursts, Plasma emission, Shock waves, Fundamental and Harmonic emissions, Solar corona}

\end{frontmatter}

%

\section{Introduction}\label{sec:intro}

Solar Radio Bursts (SRBs) are broadly classified into five classes namely type I, type II, type III, type IV and type V based on their morphology and drift rates as observed from the dynamic spectrograms (time-frequency maps). Among them type II bursts are signatures of magnetohydrodynamic (MHD) shocks that occur in the solar corona, typically triggered by transient events such as Coronal Mass Ejections (CMEs) and flares \citep{1947Natur.160..256P, Nelson1975PASA....2..370N, Mann1995, Aurass1997, Gopalswamy2006,Gopalswamy13,Nindos2011A&A...531A..31N,Ramesh12}. In dynamic radio spectra, type II bursts appear as narrow band features that drift from high to low frequencies at a rate of approximately 1~MHz~s$^{-1}$. These bursts are observed from meter wavelengths down to Deca-Hectometer (DH) and kilometer (KM) wavelengths. Type II bursts observed above 3~$R_\odot$ are generally referred to as interplanetary Type II bursts \citep{Gopalswamy2006}. 

It is known that type II bursts originate via the plasma emission mechanism. Firstly, a shock triggered by either CMEs or flare-associated blast waves excites plasma oscillations, called Langmuir oscillations, which are subsequently converted into electromagnetic waves through non-linear processes \citep{1985srph.book.....M}. Furthermore, some type II bursts exhibit both fundamental (F) and harmonic (H) structures \citep{1954AuJPh...7..439W, Ganse2012, Kontar2017, Morosan2023}. The frequency ratio between F and H emissions varies from 1.6 to 2.2, with a mean value of 1.8. In certain cases, second and third harmonic emissions have also been observed \citep{Bacchini2024}. Additionally, both F and H emissions may further split into a pair of bands, a phenomenon known as band splitting, which was first reported by \citet{1954AuJPh...7..439W}. 
Type II bursts may also exhibit discrete sources or fragmentation, likely caused by turbulence and inhomogeneities in the ambient medium \citep{2021ApJ...921....3C, 2023ApJ...943...43R}.

Observationally, it is known that the intensity densities of both the fundamental (F) and harmonic (H) emissions of type II bursts vary from burst to burst. For instance, in some cases, the F emission is stronger, while in others, the H emission dominates. In yet other cases, both have more or less the same intensity. In this article, we attempt to understand the reasons behind such intensity variations in F and H emissions.

\section{Observations}\label{sec:observation}

We used the type II solar radio bursts that are observed with e-CALLISTO spectrometers\footnote{\url{https://www.e-callisto.org/}},
a network of radio spectrometers installed and operated over different longitudes to monitor the solar radio emission continuously \citep{Benz2009, Monstein_Csillaghy_Benz_2023}. 
In addition, observations with GLOSS \citep{Kishore2015} were also used for the present work\footnote{\url{https://www.iiap.res.in/centers/gro/grids}}. The list of stations, their longitudes and latitudes, and operating frequency range
are shown in Table \ref{tab:e-callisto}.

\begin{table}
\centering
\resizebox{\textwidth}{!}{%
\begin{tabular}{ccccc}
\hline
\textbf{S.No.} & \textbf{Station} & \textbf{\shortstack{Latitude\\(deg)}} & \textbf{\shortstack{Longitude\\(deg)}} 
& \textbf{\shortstack{Frequency\\Range (MHz)}} \\
\hline
1 & ALASKA-COHOE & N60 & W151 & 10–85 \\
2 & ALASKA-HAARP & N62 & W145 & 10–85 \\
3 & Arecibo-Observatory & N18 & W66 & 15–85 \\
4 & GREENLAND & N66 & W50 & 10–110  \\
5 & BIR & N53 & W8 & 10–100  \\
6 & ALGERIA-CRAAG & N36 & E3 & 50–160 \\
7 & AUSTRIA-Krumbach & N47 & E10 & 50–450 \\
8 & GERMANY-DLR & N53 & E13 & 10–80 \\
9 & EGYPT-Alexandria & N31 & E30 & 10–90 \\
10 & INDIA-OOTY & N11 & E77 & 50–160 \\
11 & Gauribidanur-GLOSS & N14 & E78 & 40-440 \\
12 & INDIA-GAURI & N14 & E78 & 50–170 \\
13 & SSRT & N52 & E102 & 50–425 \\
14 & Australia-ASSA & S35 & E140 & 15–85  \\
\hline
\end{tabular}%
}
\caption{Locations of the spectrometers and frequency range.}
\label{tab:e-callisto}
\end{table}

We selected 58 type II bursts which has both F and H emissions from 13 CALLISTO spectrometers and GLOSS. For those events, the start time, start frequency of F and H emissions, amplitude/intensity (in dB) for F emission ($I_F$) and H emissions ($I_H$), heliographic longitude, heliographic latitude, active region number and flare class etc were obtained using the CALLISTO spectrometer observations as well as using publicly available CDAW \footnote{\url{https://cdaw.gsfc.nasa.gov/CME_list/}} and Solar Monitoring \footnote{\url{https://solarmonitor.org/}} databases and the details are tabled in Table \ref{tab:tab2}. 

\begin{figure}
  \centering
  \caption*{(a) Harmonic is stronger (\(I_H/I_F > 1\))}
  \vspace{1em} 
    \includegraphics[scale=0.4]{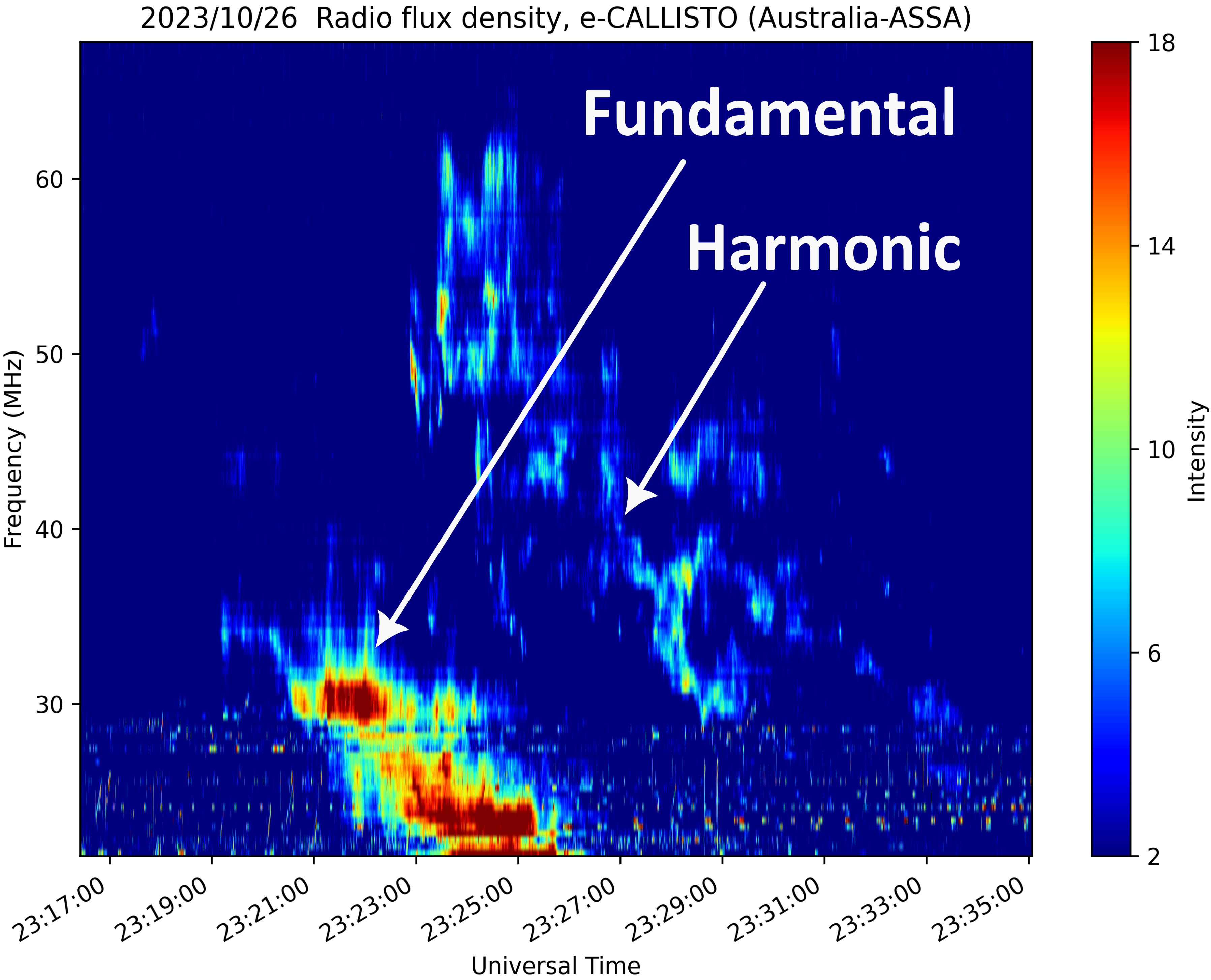}
    \includegraphics[scale=0.4]{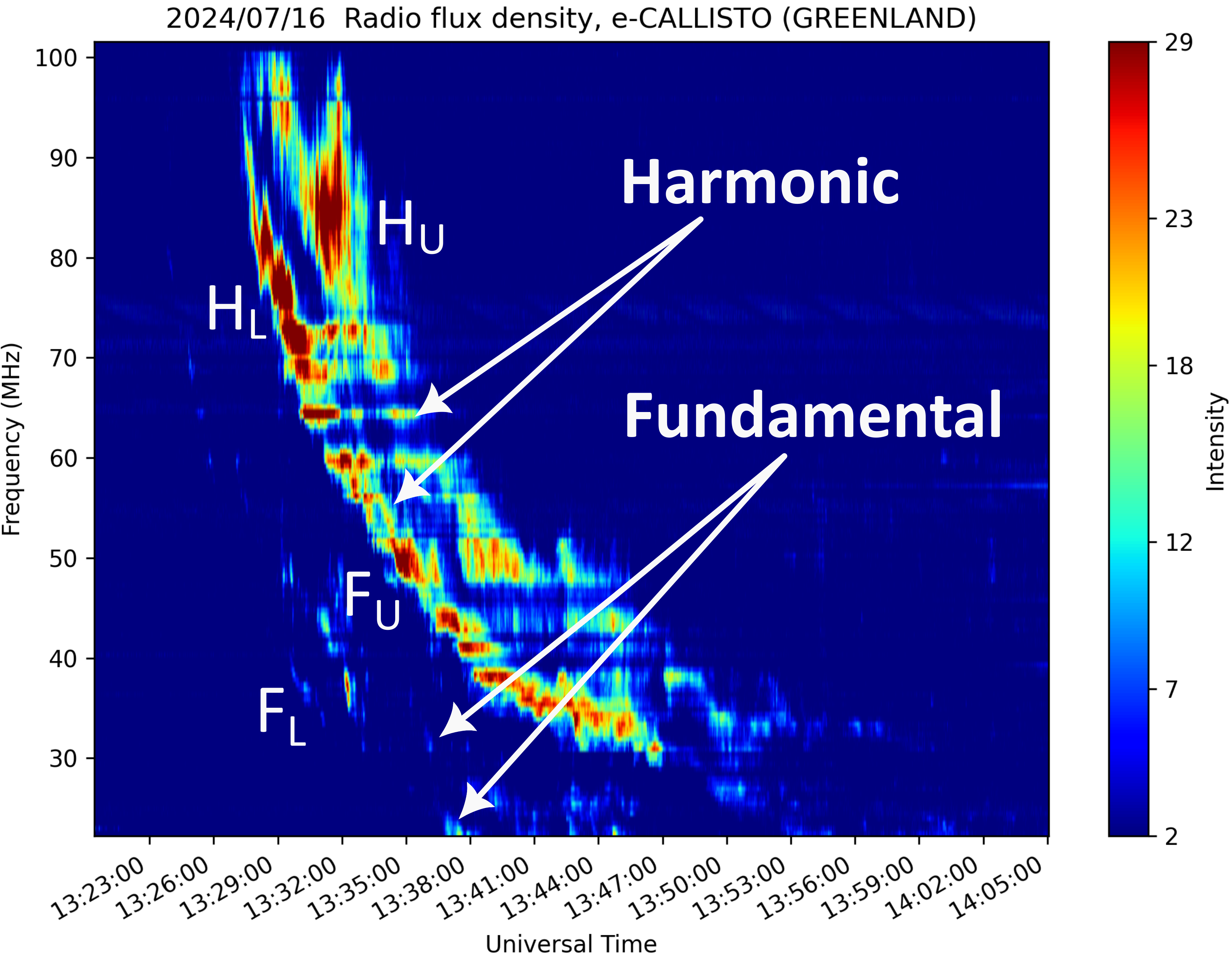}
  \caption*{(b) Fundamental is stronger (\(I_H/I_F < 1\))}
  \caption{F and H emissions of type II bursts. In upper panel F emission is stronger than the H emission. In the lower panel H emission is stronger compared to the F emission. The lower panel also shows the band splitting phenomenon. Note that the different band-split lanes are indicated as $F_U$, $F_L$, $H_U$, $H_L$.}
  \label{fig:anomaly}
\end{figure}

In Figure \ref{fig:anomaly}, upper panel shows type II burst in which fundamental emission was stronger and harmonic was weaker. This event was observed on 2023 October 26. This event occurred from an active region AR3473 located at heliographic longitude E30 and latitude N17. This event was associated with an X-ray flare of GOES class C1.4. Further in the lower panel shows another type II burst observed on 2024 July 16 using the CALLISTO spectrometer located  at Greenland station. It is evident that F emission is weaker compared to the H emission. Further it is seen clearly the band splitting of F and H emissions. This event was associated with an active region AR3738 located at heliographic longitude W85 and latitude S06. This event was associated with an X-ray flare of GOES class X1.9.
Similarly we have identified 58 type II bursts with F and H emission observed by various stations shown in Table \ref{tab:tab2}. 

\section{Results and Discussions}\label{sec:results}

In this work, we selected 58 type II bursts that exhibit both fundamental (F) and harmonic (H) emissions. We measured the intensities of both F and H emissions at a given time. These values were initially in dB scale. After converting them to a linear scale, we calculated the ratio \(I_H/I_F\). Like wise, for every event, we measured the ratio at multiple epochs in the burst and the average value is considered. 
The use of intensity ratios, particularly the average value as mentioned above, has minimized the measurement errors. Additionally, we measured various observational parameters such as start time, start frequency, drift rates, frequency and intensity ratios of the H and F emissions, heliographic coordinates, active region numbers, flare class, etc. These parameters are tabulated in Table~\ref{tab:tab2}.

{ \scriptsize 
\setlength{\tabcolsep}{1pt} 
\renewcommand{\arraystretch}{1.1} 

\begin{longtable}{cccccccccccccc}
    \caption{Details of the type II bursts observed. $f_s$ - start frequency of the burst, $V_D (= \Delta f/\Delta t$) - drift rate of the burst where $\Delta f$ is the band-width of burst and $\Delta t$ is the duration of burst, $f_H/f_F$ - frequency ratio of harmonic and fundamental emission, $I_H/I_F$ - intensity ratio of the harmonic and fundamental emission.} 
    \label{tab:tab2} \\
    
    \toprule
    Date & Station & Start & \(f_s\) & \(V_D\) &
    \(f_H/f_F\) & \(I_H/I_F\) & Location & Active & Flare \\
    & Name & Time & (MHz) & (MHz $s^{-1}$) & & & & region & class \\
    \midrule
    \endfirsthead
    
    \multicolumn{11}{l}{\small\slshape Continued from previous page} \\
    \toprule
    Date & Station & Start & \(f_s\) & \(V_D\) &
    \(f_H/f_F\) & \(I_H/I_F\) & Location & AR & Flare \\
    & Name & Time & (MHz) & (MHz/s) & & & & & class\\
    \midrule
    \endhead
    
    \midrule
    \multicolumn{11}{r}{\small\slshape Continued on next page} \\
    \endfoot
    
    \bottomrule
    \endlastfoot

13-06-2010 & Gauribidanur-GLOSS & 05:39 & 140 & 0.17 & 1.89 & 3.08 & S25W84 & AR1079 & M1.0 \\
04-09-2011 & Gauribidanur-GLOSS & 04:42 & 47 & 0.13 & 1.83 & 0.78 & N19W67 & AR1286 & C9.0 \\
02-05-2013 & Gauribidanur-GLOSS & 05:06 & 49 & 0.08 & 2.03 & 0.84 & N10W26 & AR1731 & M1.1  \\
05-11-2014 & Gauribidanur-GLOSS & 09:45 & 80 & 0.08 & 1.88 & 1.09 & N20E68 & AR2205 & M7.9 \\
09-10-2021 & Australia-ASSA &  06:34 & 71 & 0.07 & 1.82 & 0.12 &  N17E09 &  AR2882 & M1.6  \\
02-11-2021 & Australia-ASSA &  02:22 & 33 & 0.05 & 1.93 & 0.44 &  N14E01 &  AR2891 & M1.7  \\
20-12-2021 & BIR &  11:26 & 88 & 0.07 & 1.88 & 0.68 &  S22W05 &  AR2908 & M1.8  \\
12-02-2022 & INDIA-GAURI &  08:33 & 308 & 0.17 & 1.84 & 1.78 &  S19W84 &  AR2939 & M1.4  \\
02-03-2022 & GREENLAND &  17:38 & 101 & 0.10 & 1.72 & 3.76 &  N15E29 &  AR2958 & M2.0  \\
25-03-2022 & Gauribidanur-GLOSS & 05:14 & 74 & 0.20 & 2.03 & 0.41 & S21E36 & AR2974 & M1.4  \\
02-04-2022 & GREENLAND &  13:23 & 64 & 0.08 & 1.91 & 0.21 &  N15W61 &  AR2976 & M3.9  \\
17-04-2022 & INDIA-OOTY &  03:28 & 368 & 0.42 & 1.87 & 1.41 &  N12E88 &  AR2994 & X1.1  \\
30-04-2022 & Arecibo-Observatory &  13:46 & 81 & 0.09 & 1.96 & 1.26 &  N16W88 &  AR2994 & X1.1  \\
30-04-2022 & ALASKA-HAARP &  19:47 & 63 & 0.08 & 1.70 & 2.00 &  N16W88 &  AR2994 & M1.9  \\
04-07-2022 & GREENLAND &  13:32 & 86 & 0.07 & 1.90 & 0.04 &  N18E37 &  AR3050 & C5.1  \\
23-09-2022 & Arecibo-Observatory &  18:02 & 54 & 0.19 & 1.64 & 1.59 &  N19E77 &  AR3110 & M1.7  \\
29-09-2022 & Arecibo-Observatory &  12:00 & 83 & 0.05 & 1.77 & 1.12 &  N26E86 &  AR3107 & C5.7  \\
03-12-2022 & Arecibo-Observatory &  17:43 & 70 & 0.15 & 1.95 & 1.26 &  N14E89 &  AR3157 & M1.2  \\
20-01-2023 & Arecibo-Observatory &  14:01 & 85 & 0.08 & 1.78 & 0.41 &  S22W07 &  AR3190 & C5.3  \\
24-02-2023 & ALASKA-HAARP &  20:21 & 50 & 0.08 & 1.91 & 0.22 &  N29W24 &  AR3229 & M3.7  \\
03-03-2023 & Arecibo-Observatory & 18:05 & 76 & 0.02 & 1.72 & 1.13 &  N22W80 &  AR3234 & X2.1  \\
21-04-2023 & Arecibo-Observatory & 17:56 & 74 & 0.15 & 2.06 & 0.56 &  S22W11 &  AR3283 & M1.7  \\
04-05-2023 & GERMANY-DLR &  08:36 & 68 & 0.06 & 1.79 & 0.50 &  N17E43 &  AR3296 & M3.9  \\
06-05-2023 & ALASKA-COHOE &  00:47 & 54 & 0.09 & 1.62 & 0.57 &  N13E43 &  AR3297 & C2.8  \\
09-05-2023 & SSRT &  03:51 & 438 & -1.02 & 1.77 & 0.11 &  N12W13 &  AR3296 & M6.5  \\
11-05-2023 & INDIA-GAURI &  08:57 & 160 & 0.12 & 1.66 & 0.79 &  S06W41 &  AR3294 & M2.1  \\
17-05-2023 & Arecibo-Observatory &  15:22 & 65 & 0.04 & 1.96 & 0.22 &  S18W45 &  AR3309 & C4.3  \\
12-06-2023 & INDIA-OOTY &  07:02 & 82 & 0.10 & 1.93 & 0.11 &  N17W44 &  AR3330 & C5.2  \\
20-06-2023 & Arecibo-Observatory & 17:02 & 49 & 0.09 & 1.89 & 0.35 &  S17E73 &  AR3341 & X1.1  \\
02-07-2023 & SSRT &  02:34 & 211 & 0.62 & 1.67 & 0.79 &  S22E55 &  AR3359 & M2.0  \\
28-07-2023 & ALASKA-HAARP & 15:52 & 74 & 0.10 & 1.92 & 0.35 &  N27W87 &  AR3376 & M4.1  \\
17-08-2023 & GERMANY-DLR & 12:36 & 42 & 0.07 & 1.88 & 1.59 &  N19W80 &  AR3397 & C5.1  \\
26-08-2023 & GREENLAND & 12:56 & 58 & 0.07 & 1.88 & 2.51 &  N10W70 &  AR3405 & C1.9  \\
26-10-2023 & Australia-ASSA & 23:17 & 51 & 0.06 & 1.84 & 0.14 &  N17E30 &  AR3473 & C1.4  \\
29-01-2024 & Australia-ASSA & 04:11 & 85 & 0.05 & 1.84 & 1.12 &  N28W86 &  AR3559 & M6.8  \\
02-02-2024 & Australia-ASSA & 03:07 & 52 & 0.06 & 1.76 & 5.21 &  S18E60 &  AR3571 & M1.1  \\
07-02-2024 & Australia-ASSA & 03:21 & 83 & 0.03 & 2.00 & 0.11 &  S40W78 &  AR3575 & M5.1  \\
16-02-2024 & Australia-ASSA & 06:53 & 85 & 0.05 & 1.86 & 0.79 &  S16W80 &  AR3576 & X2.5  \\
10-03-2024 & ALGERIA-CRAAG & 12:10 & 151 & 0.11 & 1.98 & 0.45 &  S13W38 &  AR3599 & M7.4  \\
23-04-2024 & GREENLAND & 17:45 & 49 & 0.04 & 1.70 & 0.83 &  S08W59 &  AR3645 & M2.9  \\
08-05-2024 & Australia-ASSA & 05:09 & 33 & 0.05 & 1.80 & 0.50 &  S22W11 &  AR3664 & X1.0  \\
13-07-2024 & GREENLAND & 12:42 & 52 & 0.03 & 1.87 & 1.18 &  S08W42 &  AR3738 & M5.3  \\
16-07-2024 & GREENLAND & 13:23 & 94 & 0.06 & 1.85 & 4.48 &  S06W85 &  AR3738 & X1.9  \\
24-07-2024 & GREENLAND & 17:26 & 37 & 0.07 & 1.82 & 0.47 &  S07W71 &  AR3751 & M2.9  \\
29-07-2024 & Gauribidanur-GLOSS & 02:36 & 172 & 0.35 & 1.88 & 0.28 & S05W04 & AR3766 & X1.5 \\
05-08-2024 & GREENLAND & 15:30 & 53 & 0.05 & 1.91 & 0.20 &  S08E55 &  AR3780 & X1.1  \\
07-08-2024 & GREENLAND & 18:48 & 50 & 0.09 & 1.83 & 1.47 &  S07W07 &  AR3777 & M5.0  \\
01-10-2024 & ALASKA-COHOE & 22:17 & 73 & 0.06 & 1.90 & 0.16 &  S18E16 &  AR3842 & X7.1  \\
02-10-2024 & Australia-ASSA & 05:44 & 83 & 0.06 & 1.85 & 0.71 &  S15E20 &  AR3842 & M3.6  \\
03-10-2024 & AUSTRIA-Krumbach & 12:16 & 181 & 0.31 & 1.98 & 0.09 &  S15W03 &  AR3842 & X9.0  \\
09-10-2024 & Australia-ASSA & 01:45 & 42 & 0.09 & 1.98 & 0.22 &  N13W08 &  AR3848 & X1.8  \\
14-10-2024 & ALASKA-HAARP & 00:15 & 62 & 0.09 & 1.99 & 0.32 &  N10W78 &  AR3848 & M3.4  \\
25-11-2024 & Australia-ASSA & 20:43 & 69 & 0.10 & 1.96 & 1.26 &  N20E89 &  AR3908 & M1.1  \\
04-12-2024 & INDIA-OOTY & 10:04 & 86 & 0.12 & 1.70 & 5.01 &  S09E89 &  AR3917 & M2.3  \\
08-12-2024 & INDIA-OOTY & 09:04 & 119 & 0.12 & 1.85 & 4.47 &  S08W54 &  AR3912 & X2.2  \\
20-12-2024 & EGYPT-Alexandria & 10:13 & 140 & 0.21 & 1.71 & 0.35 &  S13E77 &  AR3932 & C9.4  \\
24-12-2024 & INDIA-OOTY & 08:41 & 149 & 0.09 & 1.98 & 0.25 &  S19E18 &  AR3932 & M4.1  \\
25-12-2024 & SSRT & 04:48 & 171 & 0.39 & 1.88 & 0.64 &  S18E06 &  AR3932 & M4.9  \\

\end{longtable}
}

\begin{figure}
    \centering
    \includegraphics[width=1.0\textwidth]{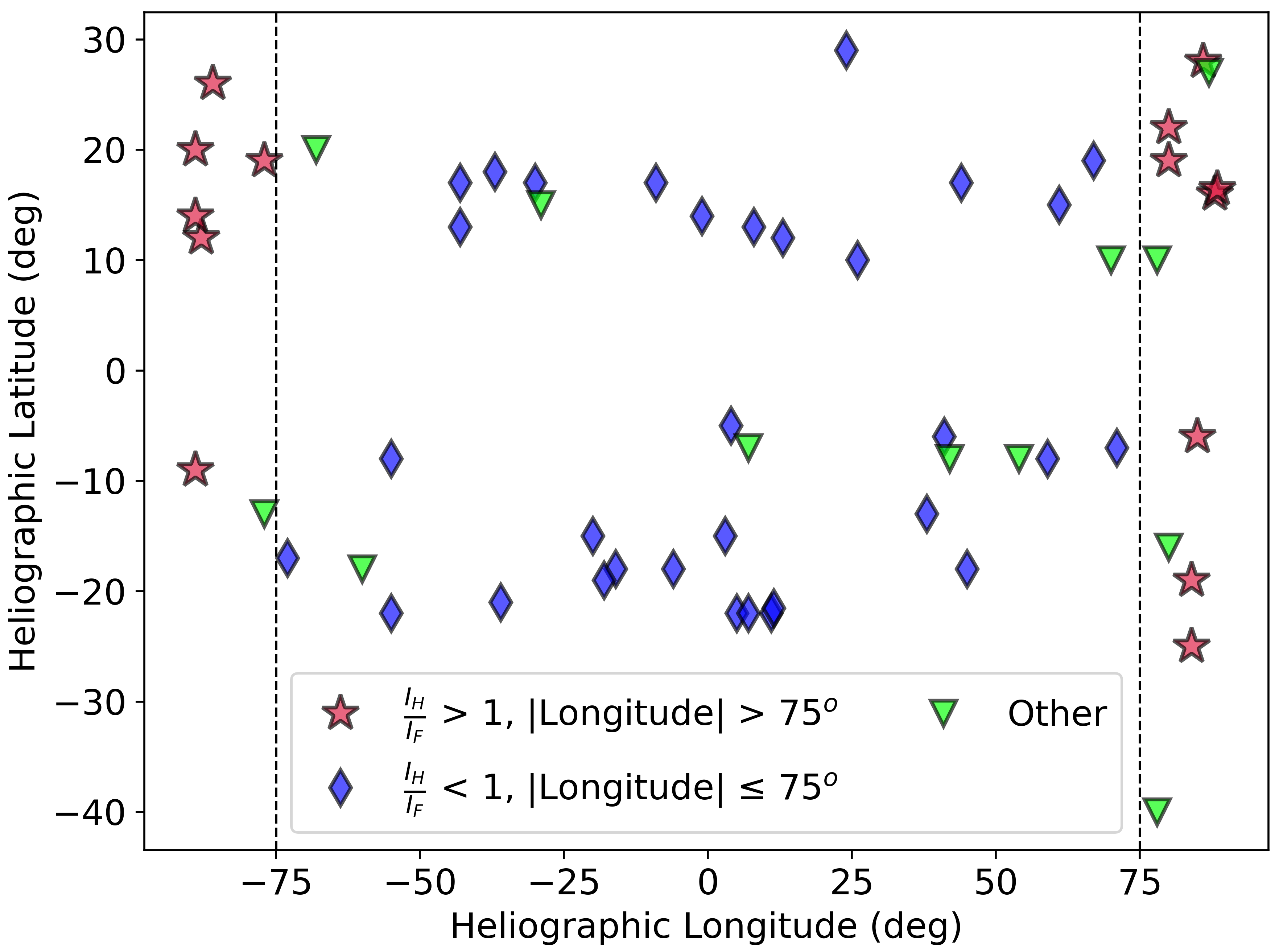}
    \caption{Location of the active regions associated with type II bursts. The blue diamonds indicate type II bursts that originated from and are associated with active regions whose longitudes are $<75^\circ$ and have $I_H/I_F < 1$. The red stars indicate type II bursts associated with active regions whose longitudes are $>75^\circ$ and have $I_H/I_F > 1$. The green diamonds indicate those that violate the conditions mentioned above.}
    \label{fig:type_II_SRB_properties}
\end{figure}

In this study, we find that 58 type II bursts exhibit a systematic dependence of the harmonic-to-fundamental intensity ratio on the heliographic longitude, as shown in Figure~\ref{fig:type_II_SRB_properties}. Among the 19 events located beyond heliographic longitudes of \(\pm75^\circ\), 14 (\(74\%\)) exhibit a harmonic-to-fundamental intensity ratio \(I_H/I_F > 1\). In contrast, among the remaining 39 events occurring within heliographic longitudes of \(\pm75^\circ\), 32 (\(82\%\)) show a intensity ratio \(I_H/I_F < 1\), suggesting dominant fundamental emission for disk events (see Table~\ref{tab:directional-type2-srbs}). There is no correlation between the flare class and the ratio \(I_H/I_F \).

According to  \citet{1974IAUS...57..239C}, ground-based observations of type~III radio bursts associated with sunspot regions located about $70^{\circ}$ to either the east or west of the solar central meridian are predominantly harmonic (H) emission. The fundamental (F) component, however, shows a limiting directivity of $\pm65^{\circ}$ from the central meridian at 80~MHz \citep{Suzuki1982, 2013ApJ...775...38S}. 
Furthermore, Morimoto et al 1963 suggests that 2/3 of intensity of type II bursts drops down for an angles between $60^\circ-90^\circ$. Therefore we took average number $75^{\circ}$ and a vertical line was drawn at $75^{\circ}$ in Figure \ref{fig:type_II_SRB_properties}.

According to \citet{1985srph.book.....M} and \citet{Nelson1975PASA....2..370N}, near the disk center the F emission usually exceeds the H emission in brightness. For events near or behind the limb the F emission is strongly attenuated by propagation effects due to density gradient in the corona.

Considering the close association between type II bursts and CMEs, we investigated the angular width of the CMEs associated with the type II bursts in our list. The mean angular width of the CMEs for the cases \(I_H/I_F > 1\) and \(I_H/I_F < 1\) are ${\approx}145^{\circ}$ and
${\approx}102^{\circ}$, respectively. The maximum angular width of the CMEs for the case \(I_H/I_F < 1\) in our study is $180^{\circ}$. 
This implies that H emission is more intense than the F emission for wider CMEs. 

Reports indicate a close association between the angular widths of the CMEs and the type II bursts. Wider CMEs have wider shocks and larger area where the electrons gain
energy via shock acceleration \citep{Michalek2007,Ramesh2022}. CMEs/shocks with smaller radius of curvature are expected to be radio-quiet \citep{Cairns2003}. These results indicate that the probability of type II bursts occurring at locations other than the nose (leading edge) of the associated MHD shock is likely to be higher in the case of the wider CMEs. This is  similar to the above mentioned discussion about type II bursts closer to the disk center and at the limb. So, it is possible that F emission can be more attenuated in the case of wider CMEs. 
It is known that radio bursts exhibit directional characteristics \citep{Mori1963, 1974A&A....37..109S, Suzuki1982, Ramesh2011, 2013ApJ...762...89R, 2020SoPh..295..153M}. Therefore, the chances of the observed variation in the intensities of the fundamental and harmonic emissions influenced by the directivity and the viewing angle too are there.

Note that directivity is defined as the ratio of the power received within a specified angular range from a source embedded in a scattering and refracting medium, to the power that would be received from the same source, at the same location and with the same total emitted power, if it were radiating isotropically in a vacuum \citep{2007ApJ...671..894T}.

The average drift rate and frequency ratio of the type II bursts in our study are ${\approx}$0.13~MHz/s and ${\approx}$1.86, respectively. These are consistent with the similar values reported for type II bursts \citep{Nelson1985srph.book..333N,Mann1995}. There is east-west asymmetry in the locations of active regions associated with the type II bursts \citep{Sel2016}. 

In our study, we find that 24 events are from the East heliographic longitudes and 34 are observed from the Western heliographic longitudes. Further, all the events except one, originated from active regions located within the heliographic latitudes 
$\pm 10^\circ - 30^\circ$.

\begin{table}
\setcounter{table}{2}
\renewcommand{\arraystretch}{1.5}
\begin{tabular}{llll}
\hline
\shortstack{\textbf{Heliographic} \\ \textbf{Longitude (L)}} &
\shortstack{\textbf{Number of} \\ \textbf{Type II SRBs}} &
$ \frac{I_H}{I_F} > 1$ &
$ \frac{I_H}{I_F} < 1$ \\
\hline
$|L| > 75^{\circ}$ & 19 & 14 & 5 \\ 
$|L| \le 75^{\circ}$ & 39 & 7 & 32 \\
\hline
\end{tabular}
\caption{\textbf{Statistics of type II bursts and their intensity ratios}}
\label{tab:directional-type2-srbs}
\end{table}





  

\section{Summary}

In this study, we explored 58 type II bursts that showed F and H emissions to understand large fraction $\approx 64\%$ of type II bursts occur with strong F emissions and $\approx 36\%$ of the bursts show strong H emissions. In some cases, both F and H emissions are nearly same. For this study we used the data obtained from CALLISTO spectrometers and GLOSS. We traced back to the activities from which type II bursts was originated using the existing catalogs mentioned previously. Our main results are:

\begin{enumerate}
    \item The bursts revealed a systematic dependence of the harmonic-to-fundamental intensity ratio of type II bursts on the heliographic longitude.
    
    \item Among the 19 events located beyond heliographic longitudes of 
\(\pm75^\circ\), 14 (\(74\%\)) exhibit a intensity ratio \(I_H/I_F > 1\). This indicates that H emissions are stronger when the active region associated with type II bursts are located at heliograhic longitudes greater than 
\(\pm75^\circ\). 
    
    \item Among 39 events occurring within \(\pm75^\circ\), 32 (\(82\%\)) show \(I_H/I_F < 1\).
     This indicates that F emissions are stronger when the active region associated with type II bursts are located at heliograhic longitudes less than \(\pm75^\circ\). 

    \item The above results suggest that the observed relative variation in the intensities of the fundamental and harmonic emissions is a consequence of refraction due to density gradient in the solar corona, directivity and viewing angle of the bursts.

    \item Among the 58 type II bursts, 12 events differ from the above pattern. 
     It is possible that the associated MHD shocks could have been non-radial for these events. 



    \end{enumerate}
    
We intend to carry out a similar study with larger data set to verify the above results.
%


\begin{acks}
We gratefully acknowledge the use of dynamic spectrogram data from the e‑CALLISTO network stations: ALASKA‑COHOE, ALASKA‑HAARP, ALGERIA‑CRAAG, Arecibo Observatory, Australia‑ASSA, AUSTRIA‑Krumbach, BIR, EGYPT‑Alexandria, GERMANY‑DLR, GREENLAND, INDIA‑GAURI, INDIA‑OOTY, and SSRT. 
We thank Indrajit V. Barve and G. V. S. Gireesh for discussions on the Gauribidanur-GLOSS data, and SolarMonitor.org team for providing details/images related to the different solar observations.
We also acknowledge the use of SOHO/LASCO CME Catalog, generated and maintained at the CDAW Data Center by NASA and The Catholic University of America in cooperation with the Naval Research Laboratory. SOHO is a project of international cooperation between ESA and NASA. We thank anonymous referee for providing constructive comments and suggestions of the manuscript.
\end{acks}

%
%
%
%
%
%
%

%
%
\bibliographystyle{spr-mp-sola}
\bibliography{main}  

@INCOLLECTION{Aurass1997,
       author = {{Aurass}, Henry},
        title = "{Coronal Mass Ejections and Type II Radio Bursts}",
    booktitle = {Coronal Physics from Radio and Space Observations},
         year = 1997,
      journal = {Lecture Notes in Physics},
       editor = {{Trottet}, Gerard},
       volume = {483},
        pages = {135},
          doi = {10.1007/BFb0106455},
       adsurl = {https://ui.adsabs.harvard.edu/abs/1997LNP...483..135A}
}

@ARTICLE{2007ApJ...671..894T,
       author = {{Thejappa}, G. and {MacDowall}, R.~J. and {Kaiser}, M.~L.},
        title = "{Monte Carlo Simulation of Directivity of Interplanetary Radio Bursts}",
      journal = {\apj},
         year = 2007,
        month = dec,
       volume = {671},
       number = {1},
        pages = {894-906},
          doi = {10.1086/522664},
       adsurl = {https://ui.adsabs.harvard.edu/abs/2007ApJ...671..894T}
}

@ARTICLE{Kontar2017,
       author = {{Kontar}, E.~P. and {Yu}, S. and {Kuznetsov}, A.~A. and {Emslie}, A.~G. and {Alcock}, B. and {Jeffrey}, N.~L.~S. and {Melnik}, V.~N. and {Bian}, N.~H. and {Subramanian}, P.},
        title = "{Imaging spectroscopy of solar radio burst fine structures}",
      journal = {Nature Communications},
         year = 2017,
        month = nov,
       volume = {8},
          eid = {1515},
        pages = {1515},
          doi = {10.1038/s41467-017-01307-8},
archivePrefix = {arXiv},
       eprint = {1708.06505},
 primaryClass = {astro-ph.SR},
       adsurl = {https://ui.adsabs.harvard.edu/abs/2017NatCo...8.1515K}
}

@ARTICLE{Benz2009,
       author = {{Benz}, A.~O. and {Monstein}, C. and {Meyer}, H. and {Manoharan}, P.~K. and {Ramesh}, R. and {Altyntsev}, A. and {Lara}, A. and {Paez}, J. and {Cho}, K. -S.},
        title = "{A World-Wide Net of Solar Radio Spectrometers: e-CALLISTO}",
      journal = {Earth Moon and Planets},
         year = 2009,
        month = apr,
       volume = {104},
       number = {1-4},
        pages = {277-285},
          doi = {10.1007/s11038-008-9267-6},
       adsurl = {https://ui.adsabs.harvard.edu/abs/2009EM&P..104..277B}
}

@ARTICLE{1947Natur.160..256P,
       author = {{Payne-Scott}, Ruby and {Yabsley}, D.~E. and {Bolton}, J.~G.},
        title = "{Relative Times of Arrival of Bursts of Solar Noise on Different Radio Frequencies}",
      journal = {\nat},
         year = 1947,
        month = aug,
       volume = {160},
       number = {4060},
        pages = {256-257},
          doi = {10.1038/160256b0},
       adsurl = {https://ui.adsabs.harvard.edu/abs/1947Natur.160..256P}
}

@ARTICLE{1985srph.book.....M,
	author = {{McLean}, D.~J. and {Labrum}, N.~R.},
	title = "{Solar radiophysics : studies of emission from the sun at metre wavelengths}",
	year = 1985,
    journal = {Cambridge University Press},
	adsurl = {https://ui.adsabs.harvard.edu/abs/1985srph.book.....M}
}

@ARTICLE{Gopalswamy2006,
       author = {{Gopalswamy}, Nat},
        title = "{Coronal Mass Ejections and Type II Radio Bursts}",
      journal = {Geophysical Monograph Series},
         year = 2006,
        month = oct,
       volume = {165},
        pages = {207},
          doi = {10.1029/9781118666203.ch18},
       adsurl = {https://ui.adsabs.harvard.edu/abs/2006GMS...165..207G}
}

@ARTICLE{Mann1995,
       author = {{Mann}, G. and {Classen}, T. and {Aurass}, H.},
        title = "{Characteristics of coronal shock waves and solar type II radio bursts.}",
      journal = {\aap},
         year = 1995,
        month = mar,
       volume = {295},
        pages = {775},
       adsurl = {https://ui.adsabs.harvard.edu/abs/1995A&A...295..775M}
}

@ARTICLE{Nelson1975PASA....2..370N,
       author = {{Nelson}, G.~J. and {Robinson}, R.~D.},
        title = "{Multi-Frequency Heliograph Observations of Type II Bursts}",
      journal = {\pasa},
         year = 1975,
        month = oct,
       volume = {2},
        pages = {370},
          doi = {10.1017/S1323358000014363},
       adsurl = {https://ui.adsabs.harvard.edu/abs/1975PASA....2..370N}
}

@ARTICLE{Nelson1985srph.book..333N,
       author = {{Nelson}, G.~J. and {Melrose}, D.~B.},
        title = "{Type II bursts.}",
    booktitle = {Solar Radiophysics: Studies of Emission from the Sun at Metre Wavelengths},
         year = 1985,
      journal = {Cambridge University Press},

       editor = {{McLean}, D.~J. and {Labrum}, N.~R.},
        pages = {333-359},
       adsurl = {https://ui.adsabs.harvard.edu/abs/1985srph.book..333N}
}

@ARTICLE{Nindos2011A&A...531A..31N,
       author = {{Nindos}, A. and {Alissandrakis}, C.~E. and {Hillaris}, A. and {Preka-Papadema}, P.},
        title = "{On the relationship of shock waves to flares and coronal mass ejections}",
      journal = {\aap},
         year = 2011,
        month = jul,
       volume = {531},
          eid = {A31},
        pages = {A31},
          doi = {10.1051/0004-6361/201116799},
archivePrefix = {arXiv},
       eprint = {1105.1268},
 primaryClass = {astro-ph.SR},
       adsurl = {https://ui.adsabs.harvard.edu/abs/2011A&A...531A..31N}
}

@ARTICLE{Ganse2012,
       author = {{Ganse}, U. and {Kilian}, P. and {Vainio}, R. and {Spanier}, F.},
        title = "{Emission of Type II Radio Bursts - Single-Beam Versus Two-Beam Scenario}",
      journal = {\solphys},
         year = 2012,
        month = oct,
       volume = {280},
       number = {2},
        pages = {551-560},
          doi = {10.1007/s11207-012-0077-7},
archivePrefix = {arXiv},
       eprint = {1206.5712},
 primaryClass = {astro-ph.SR},
       adsurl = {https://ui.adsabs.harvard.edu/abs/2012SoPh..280..551G}
}

@ARTICLE{Morosan2023,
       author = {{Morosan}, D.~E. and {Pomoell}, J. and {Kumari}, A. and {Kilpua}, E.~K.~J. and {Vainio}, R.},
        title = "{A type II solar radio burst without a coronal mass ejection}",
      journal = {\aap},
         year = 2023,
        month = jul,
       volume = {675},
          eid = {A98},
        pages = {A98},
          doi = {10.1051/0004-6361/202245515},
archivePrefix = {arXiv},
       eprint = {2305.11545},
 primaryClass = {astro-ph.SR},
       adsurl = {https://ui.adsabs.harvard.edu/abs/2023A&A...675A..98M}
}

@misc{Monstein_Csillaghy_Benz_2023, title={CALLISTO Quicklook Solar Spectrogram Plots}, url={https://spase-metadata.org/ISWI/DisplayData/Callisto/FAS/PT15M}, DOI={10.48322/WY0B-TQ35}, abstractNote={This dataset contains solar dynamic spectrogram PNG plots of the Callisto spectrometer data from the e-Callisto network of stations. Each plot spans 15 minutes.  From the website http://e-callisto.org - The CALLISTO spectrometer is a programmable heterodyne receiver built in the framework of IHY2007 and ISWI by former Radio and Plasma Physics Group (PI Christian Monstein) at ETH Zurich, Switzerland. The main applications are observation of solar radio bursts and rfi-monitoring for astronomical science, education and outreach. Many CALLISTO instruments have been deployed. A complete list of stations is available from the website.}, publisher={International Space Weather Initiative}, author={Monstein, Christian and Csillaghy, André and Benz, Arnold O.}, year={2023} }

@ARTICLE{Sel2016,
	author = {{Selvarani}, G. and {Suresh}, K. and {Shanmugaraju}, A.},
	title = "{Investigation on the source location of flares associated with type II radio bursts using multi-wavelength observations}",
	journal = {Ind.J. Radio Space Phys.},
	year = 2016,
	month = dec,
	volume = {45},
	pages = {154-159},
	url = {http://nopr.niscpr.res.in/handle/123456789/41332},
}

@ARTICLE{1954AuJPh...7..439W,
       author = {{Wild}, J.~P. and {Murray}, J.~D. and {Rowe}, W.~C.},
        title = "{Harmonics in the Spectra of Solar Radio Disturbances}",
      journal = {Aust. J. Phys.},
         year = 1954,
        month = sep,
       volume = {7},
        pages = {439},
          doi = {10.1071/PH540439},
       adsurl = {https://ui.adsabs.harvard.edu/abs/1954AuJPh...7..439W}
}

@ARTICLE{Mori1963,
       author = {{Morimoto}, M.},
        title = "{On the Directivity of Solar Radio Bursts at Meter Waves}",
      journal = {\pasj},
         year = 1963,
        month = jan,
       volume = {15},
        pages = {46},
       adsurl = {https://ui.adsabs.harvard.edu/abs/1963PASJ...15...46M}
}

@ARTICLE{2021ApJ...921....3C,
       author = {{Carley}, Eoin P. and {Cecconi}, Baptiste and {Reid}, Hamish A. and {Briand}, Carine and {Sasikumar Raja}, K. and {Masson}, Sophie and {Dorovskyy}, Vladimir and {Tiburzi}, Caterina and {Vilmer}, Nicole and {Zucca}, Pietro and {Zarka}, Philippe and {Tagger}, Michel and {Grie{\ss}meier}, Jean-Mathias and {Corbel}, St{\'e}phane and {Theureau}, Gilles and {Loh}, Alan and {Girard}, Julien N.},
        title = "{Observations of Shock Propagation through Turbulent Plasma in the Solar Corona}",
      journal = {\apj},
         year = 2021,
        month = nov,
       volume = {921},
       number = {1},
          eid = {3},
        pages = {3},
          doi = {10.3847/1538-4357/ac1acd},
archivePrefix = {arXiv},
       eprint = {2108.05587},
 primaryClass = {astro-ph.SR},
       adsurl = {https://ui.adsabs.harvard.edu/abs/2021ApJ...921....3C}
}

@ARTICLE{2013ApJ...775...38S,
       author = {{Sasikumar Raja}, K. and {Ramesh}, R.},
        title = "{Low-frequency Observations of Transient Quasi-periodic Radio Emission from the Solar Atmosphere}",
      journal = {\apj},
         year = 2013,
        month = sep,
       volume = {775},
       number = {1},
          eid = {38},
        pages = {38},
          doi = {10.1088/0004-637X/775/1/38},
archivePrefix = {arXiv},
       eprint = {1611.05227},
 primaryClass = {astro-ph.SR},
       adsurl = {https://ui.adsabs.harvard.edu/abs/2013ApJ...775...38S}
}

@INPROCEEDINGS{1974IAUS...57..239C,
       author = {{Caroubalos}, C. and {Steinberg}, J.~L.},
        title = "{Direct Measurements of the Directivity of Type i and Type III Radiation at 169 MHZ (presented by C. Caroubalos)}",
    booktitle = {Coronal Disturbances},
         year = 1974,
       editor = {{Newkirk}, Gordon Allen},
       series = {IAU Symposium},
       volume = {57},
        month = jan,
        pages = {239},
       adsurl = {https://ui.adsabs.harvard.edu/abs/1974IAUS...57..239C}
}

@ARTICLE{2023ApJ...943...43R,
       author = {{Ramesh}, R. and {Kathiravan}, C. and {Kumari}, Anshu},
        title = "{Solar Coronal Density Turbulence and Magnetic Field Strength at the Source Regions of Two Successive Metric Type II Radio Bursts}",
      journal = {\apj},
         year = 2023,
        month = jan,
       volume = {943},
       number = {1},
          eid = {43},
        pages = {43},
          doi = {10.3847/1538-4357/acaea5},
archivePrefix = {arXiv},
       eprint = {2302.00071},
 primaryClass = {astro-ph.SR},
       adsurl = {https://ui.adsabs.harvard.edu/abs/2023ApJ...943...43R}
}

@ARTICLE{2020SoPh..295..153M,
       author = {{Mahender}, Aroori and {Sasikumar Raja}, K. and {Ramesh}, R. and {Panditi}, Vemareddy and {Monstein}, Christian and {Ganji}, Yellaiah},
        title = "{A Statistical Study of Low-Frequency Solar Radio Type III Bursts}",
      journal = {\solphys},
         year = 2020,
        month = nov,
       volume = {295},
       number = {11},
          eid = {153},
        pages = {153},
          doi = {10.1007/s11207-020-01722-z},
archivePrefix = {arXiv},
       eprint = {2009.05755},
 primaryClass = {astro-ph.SR},
       adsurl = {https://ui.adsabs.harvard.edu/abs/2020SoPh..295..153M}
}

@ARTICLE{2013ApJ...762...89R,
       author = {{Ramesh}, R. and {Sasikumar Raja}, K. and {Kathiravan}, C. and {Narayanan}, A. Satya},
        title = "{Low-frequency Radio Observations of Picoflare Category Energy Releases in the Solar Atmosphere}",
      journal = {\apj},
         year = 2013,
        month = jan,
       volume = {762},
       number = {2},
          eid = {89},
        pages = {89},
          doi = {10.1088/0004-637X/762/2/89},
       adsurl = {https://ui.adsabs.harvard.edu/abs/2013ApJ...762...89R}
}

@ARTICLE{1974A&A....37..109S,
       author = {{Steinberg}, J.~L. and {Caroubalos}, C. and {Bougeret}, J.~L.},
        title = "{STEREO-1 measurements of the beam pattern of 169 MHz type I bursts on November 18, 1971.}",
      journal = {\aap},
         year = 1974,
        month = dec,
       volume = {37},
       number = {1},
        pages = {109-115},
       adsurl = {https://ui.adsabs.harvard.edu/abs/1974A&A....37..109S}
}

@ARTICLE{Kishore2015,
   author = {{Kishore}, P. and {Ramesh}, R. and {Kathiravan}, C. and {Rajalingam}, M.},
    title = "{A Low-Frequency Radio Spectropolarimeter for Observations of the Solar Corona}",
  journal = {\solphys},
     year = {2015},
   volume = 290,
    pages = {2409-2422},
    doi = {10.1007/s11207-015-0705-0},
    adsurl = {https://ui.adsabs.harvard.edu/abs/2015SoPh..290.2409K/abstract}
}

@ARTICLE{Ramesh2011,
   author = {{Ramesh}, R. and {Kathiravan}, C. and {Satya\,Narayanan}, A.},
    title = "{Low-frequency Observations of Polarized Emission from Long-lived Non-thermal Radio Sources in the Solar Corona}",
  journal = {\apj},
     year = {2011},
   volume = 734,
    pages = {39},
    doi = {10.1088/0004-637X/734/1/39},
    adsurl = {https://ui.adsabs.harvard.edu/abs/2011ApJ...734...39R/abstract}
}

@ARTICLE{Suzuki1982,
   author = {{Suzuki}, S. and {Sheridan}, K. V.},
    title = "{On the fundamental and harmonic components of low-frequency Type III solar radio bursts}",
  journal = {\pasa},
     year = {1982},
   volume = 4,
    pages = {382},
    doi = {10.1017/S1323358000021214},
    adsurl = {https://ui.adsabs.harvard.edu/abs/1982PASA....4..382S/abstract}
}

@ARTICLE{Ramesh12,
   author = {{Ramesh}, R. and {Kathiravan}, C. and {Anna Lakshmi}, M. and {Gopalswamy}, N. and 
{Umapathy}, S.},
    title = "{The Location of Solar Metric Type II Radio Bursts with Respect to the Associated Coronal Mass Ejections}",
  journal = {\apj},
     year = {2012},
   volume = 752,
    pages = {107},
    doi = {10.1088/0004-637X/752/2/107},
   adsurl = {https://ui.adsabs.harvard.edu/abs/2012ApJ...752..107R/abstract}
}

@ARTICLE{Gopalswamy13,
   author = {{Gopalswamy}, N. and {Xie}, H. and {M{\"a}kel{\"a}}, P. and {Yashiro}, S. and {Akiyama}, S. and {et al.}},
    title = "{Height of shock formation in the solar corona inferred from observations of type II radio bursts and coronal mass ejections}",
  journal = {Adv. Space Res.},
     year = {2013},
   volume = 51,
    pages = {1981-1989},
    doi = {10.1016/j.asr.2013.01.006}, 
    adsurl = {https://ui.adsabs.harvard.edu/abs/2013AdSpR..51.1981G/abstract}
}

@ARTICLE{Michalek2007,
   author = {{Michalek}, G. and {Gopalswamy}, N. and {Xie}, H.},
    title = "{Width of Radio-Loud and Radio-Quiet CMEs}",
  journal = {\solphys},
     year = {2007},
    month = dec,
   volume = 246,
    pages = {409},
    doi = {10.1007/s11207-007-9062-y}, 
    adsurl = {https://ui.adsabs.harvard.edu/abs/2007SoPh..246..409M/abstract}
}

@ARTICLE{Cairns2003,
   author = {{Cairns}, I. H. and {Knock}, S. A. and {Robinson}, P. A. and {Kuncic}, Z.},
    title = "{Type II Solar Radio Bursts: Theory and Space Weather Implications}",
  journal = {\ssr},
     year = {2003},
    month = apr,
   volume = 107,
    pages = {27},
    doi = {10.1023/A:1025503201687}, 
    adsurl = {https://ui.adsabs.harvard.edu/abs/2003SSRv..107...27C/abstract}
}

@ARTICLE{Ramesh2022,
   author = {{Ramesh}, R. and {Kathiravan}, C.},
    title = "{New Results from the Spectral Observations of Solar Coronal Type II Radio Bursts}",
  journal = {\apj},
     year = {2022},
    month = feb,
   volume = 926,
    pages = {38},
    doi = {10.3847/1538-4357/ac4bd6}, 
    adsurl = {https://ui.adsabs.harvard.edu/abs/2022ApJ...926...38R/abstract}
}

@ARTICLE{Bacchini2024,
       author = {{Bacchini}, Fabio and {Philippov}, Alexander A.},
        title = "{Fundamental, harmonic, and third-harmonic plasma emission from beam-plasma instabilities: a first-principles precursor for astrophysical radio bursts}",
      journal = {\mnras},
         year = 2024,
        month = mar,
       volume = {529},
       number = {1},
        pages = {169-177},
          doi = {10.1093/mnras/stae521},
archivePrefix = {arXiv},
       eprint = {2402.11011},
 primaryClass = {astro-ph.SR},
       adsurl = {https://ui.adsabs.harvard.edu/abs/2024MNRAS.529..169B}
}
%
%
%
%

\end{document}